\documentclass[10pt,twocolumn]{revtex4}%

\usepackage{amsfonts}
\usepackage{amsmath}
\usepackage{amssymb}
\usepackage{array}
\usepackage{graphicx}
\usepackage{amsmath}
\usepackage{amssymb}
\usepackage{stmaryrd}
\usepackage{marvosym}
\usepackage{hhline,float}
\usepackage{url}
\usepackage{hyperref}
\usepackage{natbib}
\usepackage{stmaryrd}
\usepackage{marvosym}
\usepackage{url}
\usepackage{setspace}
\usepackage[footnotesize,sl,nooneline]{caption}%
\providecommand{\U}[1]{\protect\rule{.1in}{.1in}}

\newcommand{\hide}[1]{\relax}

\newcommand{\Og}{\ensuremath{\Omega}}
\newcommand{\bD}{\ensuremath{\bar \Delta}}

\newcommand{\dwdx}{\ensuremath{g_0}}
\newcommand{\etac}{\ensuremath{\eta_{\mathrm{c}}}}
\newcommand{\ba}{\ensuremath{\bar a}}

\newcommand{\thom}{\ensuremath{t_\mathrm{hom}}}

\renewcommand{\small}[1][]{#1}
\newcommand{\si}{appendix}

\begin{document}

\title{Optomechanically induced transparency}
\author{S.\ Weis$^{1,2,\dagger}$, R.\ Rivi\`ere$^{2,\dagger}$, S.\ Del\'eglise$^{1,2,\dagger}$, E.\ Gavartin$^{1}$, O.\ Arcizet$^{3}$, A.\ Schliesser$^{1,2}$, T.\ J.\ Kippenberg$^{1}$}
\email{tobias.kippenberg@epfl.ch}
\affiliation{$^{1}$Ecole Polytechnique F$\acute{e}$d$\acute{e}$rale de Lausanne, EPFL,
1015 Lausanne, Switzerland,}
\affiliation{$^{2}$Max-Planck-Institut f{\"u}r Quantenoptik, Hans-Kopfermann-Str. 1, 85748
Garching, Germany,}
\affiliation{$^{3}$Institut N\'e\'el, 25 rue des Martyrs, 38042 Grenoble, France}
\affiliation{$^{\dagger}$These authors contributed equally to this work.}

\maketitle

{\textbf{\noindent %
Coherent interaction of laser radiation with multilevel atoms and molecules can lead to quantum interference in the electronic excitation pathways \cite{Fleischhauer2005}.
A prominent example observed in atomic three-level-systems is the phenomenon of electromagnetically induced transparency (EIT), in which a control laser induces a narrow spectral transparency window for a weak probe laser beam \cite{Harris1997}.
The concomitant rapid variation of the refractive index in this spectral window can give rise to dramatic reduction of the group velocity of a propagating pulse of probe light \cite{Kasapi1995, Hau1999}.
Dynamic control of EIT via the control laser enables even a complete stop, that is, storage, of probe light pulses in the atomic medium  \cite{Phillips2001,Liu2001}. 
Here, we demonstrate \emph{optomechanically induced transparency} (OMIT)--formally equivalent to EIT--in a cavity optomechanical system operating in the resolved sideband regime \cite{Schliesser2009, Schliesser2010, Agarwal2010}.
A control laser tuned to the lower motional sideband of the cavity resonance induces a dipole-like
interaction of optical and mechanical degrees of freedom \cite{Zhang2003, Marquardt2007, Wilson-Rae2007}.
Under these conditions, the destructive interference of excitation pathways for an intracavity probe field gives rise to a window of transparency when a two-photon resonance condition is met. 
As a salient feature of EIT, the power of the control laser determines the width and depth of the probe transparency window.
OMIT could therefore provide a new approach for delaying, slowing and storing light pulses  \cite{Schliesser2009, Schliesser2010, Chang2010} in long-lived mechanical excitations of optomechanical systems, whose optical and mechanical properties can be tailored in almost arbitrary ways in the micro- \cite{Arcizet2006a, Anetsberger2008, Groblacher2009a} and nano-optomechanical \cite{Thomson2007,Regal2008, Eichenfield2009a, Eichenfield2009, Anetsberger2009}  platforms developed to date \cite{Kippenberg2008}.
 }}

When the generic EIT effect has first been observed in an atomic gas \cite{Boller1991}, its great application potential in non-linear optics and optical (quantum) information processing was quickly recognized. 
In this context, the experimental demonstration of slowing and stopping light \cite{ Kasapi1995, Hau1999,Phillips2001,Liu2001} has particularly attracted researchers' attention, as it provides a route to implement a photonic quantum memory \cite{Lukin2001} or a classical optical buffer.
EIT has subsequently been studied in a wide variety of atomic media, but also in a number of 
solid-state systems \cite{Turukhin2002, Hemmer2001, Brunner2009} with a well-suited level structure.
More recently, a wider class of physical systems, which can be more flexibly engineered---such as coupled plasmonic \cite{Liu2009} or optical resonators \cite{Xu2006, Totsuka2007}---have been reported to display a related effect often referred to as coupled-resonantor induced transparency (CRIT) \cite{Smith2004}.
While CRIT bears some conceptual analogies to atomic EIT, the coupling of the involved resonators and the resulting transparency is usually not induced by an electromagnetic field, but 
is rather determined by the geometry of the structure, and therefore more difficult to tune. 

Recent experiments with optomechanical systems have demonstrated that the \emph{mechanical} response to thermal forces can be controlled by an optical field.
This effect has been exploited, for example, to implement optomechanical laser cooling and amplification \cite{Kippenberg2005, Schliesser2006, Arcizet2006a, Gigan2006}, strong optomechanical coupling \cite{Groblacher2009a}, and optical spring-tuning of coupled mechanical resonances \cite{Lin2010}.
It was recently suggested \cite{Schliesser2009, Schliesser2010} to employ optomechanical coupling to control the system's \emph{optical} response to a weak `probe' laser by a second, `control' laser driving the lower motional sideband.
As pointed out also by an independent study \cite{Agarwal2010}, this effect can be considered a strict optomechanical analog of EIT.
Advantageously, this form of induced transparency does not rely on naturally occuring resonances and could therefore also be applied to previously inaccesible wavelength regions such as the technologically important near-infrared.
Furthermore, already a single optomechanical element can achieve unity contrast, which in the atomic case is only possible within the setting of cavity QED \cite{Mucke2010}.
In this letter we present the first experimental observation of this effect.

The basic idea of our experiment is illustrated in figure 1.
It consists of a canonical optomechanical system that features linear optomechanical coupling in the sense that the displacement $x$ of a mechanical mode shifts the cavity resonance frequency  to a new frequency $\omega_\mathrm{c}'=\omega_{\mathrm{c}}+g_0 x$ with $g_0\equiv d\omega'_\mathrm{c}/dx$.
A control laser (frequency $\omega_{\mathrm{l}}$) maintains a control field $\bar a e^{-i \omega_\mathrm{l}t}$, containing $|\bar a|^2$ photons, in the cavity.
The static radiation pressure originating from this field displaces the mechanical mode by $\bar x$, leading to an effective detuning from the cavity resonance given by  $\bar \Delta=\omega_{\mathrm{l}}-(\omega_{\mathrm{c}}+g_0 \bar x)$.
Here we  consider the situation where the control laser is tuned close to the lower motional sideband, i.e. $\bar \Delta \approx -\Omega_\mathrm{m}$.
A second,  weak laser oscillating at $\omega_{\mathrm{p}}=\omega_{\mathrm{l}} +\Omega$, is subsequently used to probe the (modified) cavity resonance by driving an intracavity probe field contained in a perturbation term $\delta a (t)$.

In the case of a weak probe field (compared to the control field), it is straightforward to treat this scenario  by linearizing the optomechanical dynamics  \cite{Fabre1994, Mancini1994} for the mechanical displacement $x(t) = \bar x + \delta x(t)$ and the intracavity field $a(t)=(\bar a + \delta a(t))e^{-i\omega_\mathrm{l} t}$ around the steady state values  $(\bar x, \bar a)$. For the probe power transmission---that is, the ratio of the probe power returned from the system divided by the input probe power---the general expression 
\begin{align}
|t_\mathrm{p}
|^2&=\left|
1-\frac{1+i f(\Og)}
        {-i(\bD+\Og)+\kappa/2+2\bD f(\Og)} \etac \kappa
\right|^2 
\intertext{with}
  f(\Og)&=\hbar \dwdx^2 \ba^2\,\frac{\chi(\Og)}{i(\bD-\Og)+\kappa/2}.
\end{align}
can  be derived (see refs. \cite{Schliesser2009, Schliesser2010, Agarwal2010} and the \si).
Here,  $\chi(\Omega)=(m_{\mathrm{eff}}(\Omega_\mathrm{m}^2-\Omega^2-i \Gamma_\mathrm{m} \Omega))^{-1}$ is the susceptibility of the mechanical oscillator of effective mass $m_{\mathrm{eff}}$, resonance frequency $\Omega_\mathrm{m}$ and damping rate $\Gamma_\mathrm{m}$.
The optical mode is characterized by a total loss rate $\kappa=\kappa_0+\kappa_\mathrm{ex}$ and the cavity coupling parameter $\eta_{\mathrm{c}}=\kappa_\mathrm{ex}/(\kappa_\mathrm{0}+\kappa_\mathrm{ex})$.
As shown in figure 1, the presence of a control field $\bar a$ induces a transmission window for the probe beam when the resonance condition $\Omega \approx \Omega_{\mathrm{m}} $ is met.
The depth and the width of this transmission window are tunable by the power of the control beam as expected from the analogy with EIT.

In order to gain more \emph{physical} insight into how this effect arises, it is instructive to consider the occurring processes in a sideband picture.
The simultaneous presence of control and probe fields generates a radiation-pressure force oscillating at the frequency difference $\Omega$ of the two fields.
If this driving force oscillates close to the mechanical resonance frequency  $\Omega_\mathrm{m}$, the mechanical mode starts to oscillate coherently, $\delta x(t)=2\mathrm{Re}[X \,e^{-i \Omega t}]$, where $X$ denotes the (complex) oscillation amplitude.
This in turn gives rise to Stokes- and anti-Stokes scattering of light from the strong intracavity control field.
If the system resides deep enough in the resolved-sideband  (RSB) regime with $\kappa\ll\Omega_{\mathrm{m}}$, Stokes scattering  (to the optical frequency $\omega_\mathrm{l}-\Omega$) is strongly suppressed since it is highly off-resonant with the optical cavity.
We can therefore assume that only an anti-Stokes field builds up inside the cavity, $\delta a(t)\approx A^- e^{-i \Omega t}$.
However, the anti-Stokes scattered light exhibits the frequency $\omega_\mathrm{p}=\omega_\mathrm{l}+\Omega$; it is degenerate with the near-resonant probe field sent to the cavity.
Destructive interference of these two driving waves can suppress the build-up of an intracavity probe field.
Mathematically, these processes are captured by the Langevin equations of motion for the complex amplitudes $A^-$ and $X$, which require in the steady state 
\begin{align}
\left(-i \Delta^{\prime }+\kappa/2\right)\,  A^-  &=-i g_0  \bar a  X +\sqrt{\eta_\mathrm{c}  \kappa}\, \delta s_\mathrm{in} \\
{2m_\text{eff} \Omega_\mathrm{m} } (-i \Delta^{\prime }+ \Gamma_\mathrm{m}/2) \,  X &= -i{\hbar g_0  \bar a } A^-, 
\end{align}
where  we have assumed a high-quality factor of the mechanical oscillator and the control beam detuning $\bar \Delta=-\Omega_\mathrm{m}$. 
The amplitude of the probing field launched into the cavity is denoted as $\delta s_\mathrm{in}$ and we abbreviate  $\Delta^{\prime }\equiv\Omega -\Omega_\mathrm{m} $  (see \si). 

The solution for the probe field
\begin{equation}
A^- =\frac{\sqrt{\eta_\mathrm{c}  \kappa}}{(- i \Delta^{\prime}+\kappa/2)+\frac{\Omega_\mathrm{c}^2/4}{-i \Delta^{\prime}+\Gamma_\mathrm{m}/2 }}\delta  s_\mathrm{in}
\label{e:omitrsb}
\end{equation}
is of a form well-known from the response of an EIT medium to a probe field \cite{Milonni2005}.
Indeed, the coherence between the two ground states of an atomic $\Lambda$ system, and the coherence between the levels probed by the probe laser undergo the very same evolution as do the mechanical oscillation amplitude and the intracavity probe field in the case of OMIT. 
The role of the control laser's Rabi frequency in an atomic system is taken by the optomechanical coupling rate  $\Omega_\mathrm{c}=2 \bar a g_0 x_\mathrm{zpf}$, where $x_{\mathrm{zpf}}=\sqrt{\hbar/2 m_\mathrm{eff}\Omega_\mathrm{m}}$ designates the zero point spread of the mechanical oscillator.
For $\Omega_\mathrm{c}>\Gamma_\mathrm{m}, \kappa$, the system enters the strong coupling regime \cite{Marquardt2007, Dobrindt2008} investigated recently in the mechanical domain \cite{Groblacher2009a}, in which the optical and mechanical systems are hybridized to dressed states which differ by $\hbar \Omega_\mathrm{c}$ in  their energy.
Only the static radiation pressure bistability sets an upper limit for the coupling rate, in the resolved sideband regime ($\bar\Delta=-\Omega_\mathrm{m}\gg\kappa$) it necessitates $\Omega_\mathrm{c}< \Omega_\mathrm{m}$  \cite{Marquardt2007, Dobrindt2008}. 

To realize optomechanically induced transparency we employ toroidal whispering-gallery-mode microresonators shown in figure 2 as optomechanical system of choice \cite{Kippenberg2005, Schliesser2010}.
These resonators feature a unique combination of low effective mass $m_\mathrm{eff}$ and large coupling $g_0$, and can be engineered to display very low mechanical dissipation $\Gamma_\mathrm{m}$ by decoupling the mechanical radial-breathing mode (RBM) from other mechanical modes \cite{Anetsberger2008}.
The combination of high mechanical  frequency $\Omega_\mathrm{m}$ and low optical dissipation $\kappa$ furthermore allows reaching the resolved-sideband regime \cite{Schliesser2008} as discussed above.
The parameters of the device used in the experiments presented here are given by
 $(m_\mathrm{eff}, g_0/2 \pi, \Gamma_\mathrm{m}/2 \pi, \Omega_\mathrm{m}/2 \pi,\kappa/2 \pi)\approx(20\, \mathrm{ng},-12\,\mathrm{GHz}/ \mathrm{nm},41 \, \mathrm{kHz}, 51.8\, \mathrm{MHz}, 15 \, \mathrm{MHz})$, placed well in the resolved-sideband limit \cite{Schliesser2008}.
If the probe laser is scanned through the cavity resonance in the absence of the control laser, typically a simple Lorentzian exctinction dip is observed (cf.\ figure 2).
 We operate the cavity in the undercoupled regime, which together with modal coupling between counterpropagating modes (see \si) leads to a non-zero probe (amplitude) transmission $t_r=t_p(\Delta'=0,\Omega_\mathrm{c}=0)$ at  resonance even in the absence of the control beam.
 In the case of the present device, $|t_r|^2\approx 0.5$ (note that $|t_\mathrm{r}|^2<0.01$  can be achieved with silica toroids  \cite{Cai2000}).
 To separate the effects of this residual transmission from OMIT, we introduce the normalized  transmission of the probe $t_\mathrm{p}'=(t_\mathrm{p}-t_\mathrm{r})/(1-t_\mathrm{r})$.

As shown in figure 2, the experiment was carried out at cryogenic temperatures using a Helium-3 buffer gas cryostat. 
In addition to reducing the thermal Brownian motion of the mechanical oscillator this allows eliminating thermo-optical nonlinearities  \cite{Arcizet2009a} which can impede driving the lower motional sideband with a strong control laser.
The sample is mounted on a cryogenic head, which allows approaching a tapered fiber for near field evanescent coupling using piezoelectric positioners.
While an external-cavity diode laser was used for initial characterization, a low-noise, continuous-wave Titanium Sapphire operating at a wavelength of $\lambda\approx775\,\mathrm{nm}$ is employed for the actual OMIT experiments.
The Ti:sapphire laser's linewidth is reduced below $30\, \mathrm{kHz}$ by stabilization to a temperature-controlled reference cavity using the Pound-Drever-Hall technique.
This approach furthermore proved to provide sufficient mutual frequency stability of the cryogenic microresonator and the cavity-stabilized laser on the relevant scale of the cavity linewidth $\kappa$.

To detect the mechanical motion at low temperatures we employ a balanced homodyne detection scheme measuring the phase quadrature of the field emerging from the cavity \cite{Schliesser2009a}. 
This allows extracting the resonance frequency $\Omega_\mathrm{m}/2\pi$, quality factor $Q_\mathrm{m}$ and effective mass $m_{\mathrm{eff}}$ of the mechanical modes of interest. 
As in our previous work, we focus on the radial breathing mode of the resonator.
The mechanical modes' properties at low temperatures have been detailed in prior work.
They feature a complex temperature-dependence of the mechanical resonance frequencies and damping, which agrees excellently with a theoretical model taking into account that mechanical modes can couple to a bath of mechanical two level systems (TLS) via their strain field \cite{Arcizet2009a}. 
For the present toroid micro-resonators thermalized to temperatures in the range of $0.8 \,\mathrm{K}$ ($4 \,\mathrm{K}$) this results in expected mechanical Q-factor of approximately $Q_\mathrm{m}\approx10000$ $(Q_\mathrm{m}\approx1500)$.
The measured value of $Q_\mathrm{m}=51.8\,\mathrm{MHz}/41\,\mathrm{kHz}\approx 1300$ at the operating temperature of $3.78 \,\mathrm{K}$ agrees well with this prediction.

 {
To probe the cavity absorption spectrum in the presence of a control beam, we induce a frequency-tunable modulation sideband on the Ti:sapphire control laser.
We found that the most efficient way to create a sideband tunable over a wide radio-frequency span ($\gg50 \,\mathrm{MHz}$) consists in using a broadband phase modulator driven at the modulation frequency $\Omega$.
The laser light sent to the experiment thus consists of the carrier (which is used as control field at a frequency $\omega_\mathrm{l}$),  the probe beam at the frequency $\omega_\mathrm{p}=\omega_\mathrm{l}+\Omega$, and an additional field at $\omega_\mathrm{l}-\Omega$.
Keeping the laser detuned to the lower motional sideband of the cavity ($\bar \Delta\approx -\Omega_\mathrm{m}$), a sweep of the modulation frequency $\Omega$ scans the probe field through the cavity resonance.
In the RSB regime, the \emph{lower} laser sideband induced by phase modulation (at the frequency $\omega_\mathrm{l}-\Omega$) is far detuned (by $|\bar \Delta-\Omega|\gg\kappa$) from the cavity resonance in this situation and does not significantly interact with the optomechanical system.
It does however play a role in the homodyne detection scheme used. 
As shown in detail in the \si, demodulation of the total homodyne signal at the modulation frequency $\Omega$ using a network analyzer (NA) allows extracting a `transmission' homodyne signal $\thom$, which, in the RSB regime, is related to the probe transmission by the simple relation $\thom\approx 1-t_\mathrm{p}$.

Figure 3a) shows the theoretically expected response of the optomechanical system and the detected signals due to the combined presence of a control field (tuned to $\bar \Delta=-\Omega_\mathrm{m}$) and a frequency-swept  probing field.
Clearly, the OMIT dip is apparent in the intracavity probe power as described by equation (\ref{e:omitrsb}).
It occurs simultaneously with the onset of radiation-pressure-driven mechanical oscillations, as expected from our model.
The excitation of the intracavity probe field therefore is suppressed, and the transmitted field nearly equals the probe field sent to the cavity.
The lowest panel shows the homodyne signal expected in such a situation, and the five panels in 3b) show experimentally measured homodyne traces for the detunings $\bar \Delta/2\pi\in \{
-69.1\,\mathrm{MHz},
-57.6\,\mathrm{MHz},
-51.8\,\mathrm{MHz},
-44.6\,\mathrm{MHz},
-35.4\,\mathrm{MHz}
 \}$, and a control laser power of $0.5\,\mathrm{mW}$.
For different detunings of the control field $\bar \Delta$, the center of the probe response to the optical cavity occurs for the modulation frequencies  $\Omega\approx -\bar \Delta$, since the probe laser then matches the cavity resonance ($\omega_\mathrm{p}\approx \omega_\mathrm{c}$).
Importantly however, the sharp OMIT window occurs only when the two-photon resonance condition $\Omega=\Omega_\mathrm{m}$ (with $\Omega=\omega_\mathrm{p}-\omega_\mathrm{l}$) is met, independent of the detuning $\bar \Delta$ of the control beam---giving clear evidence to the theoretically suggested underlying mechanism.

To analyze the effect of the control beam more systematically, its detuning was fixed to the lower motional sideband. 
Varying its power from $0.125$ to $6.5\,\mathrm{ mW}$, traces of the homodyne signal are taken in the vicinity of the two-photon resonance (figure 4).
Dips of increasing depth and width are observed, which can be modeled by a simple Lorentzian function.
The minimum homodyne signal  is obtained under the condition of the two-photon resonance $\Delta'=0$.
In this case, the homodyne signal power and the probe power transmission are simply interrelated by $ |t'_\mathrm{p}|^2= \left(1- |\thom'|\right)^2$, where $\thom'=\thom/(1-t_\mathrm{r})$ is the normalized homodyne signal.
From the model (\ref{e:omitrsb}), the expected probe transmission on resonance is simply given by
\begin{equation}
  t_\mathrm{p}'(\Delta'=0)=\frac{C}{C+1},
\end{equation}
where $C\equiv\Omega_\mathrm{c}^2/\Gamma_\mathrm{m} \kappa$ is an equivalent optomechanical cooperativity parameter.
Our data match the expected curve very well if we allow for a linear correction factor in the optomechanical coupling frequency $\Omega_\mathrm{c}$ due to modal coupling and taper losses in the cryostat (see \si).
We have reached probe power transmission $|t'_\mathrm{p}|^2$ up to $81 \%$, indicating the high contrast which can be achieved in OMIT.
Higher values of $C$ could already be achieved by cooling the device to a lower temperature.

The simple relation between the homodyne signal and the probe transmission furthermore implies that the width $\Gamma_\mathrm{OMIT}$ of the coupling-induced transmission window in $|t_p'|^2$ equals the width of the measured dip in the normalized homodyne signal $|\thom'|^2$.
The values extracted from our data are shown in figure 4c along with the expected behaviour
\begin{equation}
  \Gamma_{\mathrm{OMIT}} \approx \Gamma_\mathrm{m} (1+C),
\end{equation}
with transparency windows wider than $500\,\mathrm{kHz}$ achieved in our experiment.

Concerning the implications of this work, we note that in any optomechanical system reaching a cooperativity parameter $C$ of order unity, the probe transmission can be significantly altered by the control beam, as desired, for example, in all-optical switches. 
Interestingly, the systems available already today, reach $C\approx 1$ with only thousands \cite{Anetsberger2009} or even hundreds \cite{Schliesser2008, Regal2008} of control photons in the cavity, and recently emerging integrated nano-optomechanical structures \cite{Eichenfield2009a} may be able to further reduce this number.
The resulting extreme optical non-linearities could be of interest for both fundamental and applied studies.

The tunable probe transmission window by necessity also modifies the propagation dynamics of a probe pulse sent to the optomechanical system due to the variation of the complex phase picked by its different frequency components.
Indeed, a probe pulse centered at the (shifted) cavity resonance frequency $\omega_\mathrm{c}+g_0 \bar x$ experiences a group delay of $\tau_\mathrm{g}\approx 2/\Gamma_\mathrm{OMIT}$ in the regime $C\gtrsim 1$ of interest (see \si).  
Group delays up to $\Gamma_\mathrm{m}^{-1}$ can therefore be achieved, exceeding times of several seconds in some available optomechanical systems \cite{Thomson2007}.
However, undistorted pulse propagation only occurs if the full probe pulse spectrum is contained within the transparency window of the system.
This restricts the effectivity of such a delay due to the fixed delay-bandwidth product of  $\tau_\mathrm{g} \Gamma_\mathrm{OMIT}\approx2$.
A cascade of systems may alleviate this shortcoming---the most interesting scenario being a large array of concatenated optomechanical systems.
The group delay could then be dynamically tuned while the probe pulse is propagating through the array  \cite{Schliesser2009, Schliesser2010, Chang2010}.
Such systems are closely related to an array of coupled \emph{optical} resonators, for which the possibility of light storage has been derived previously \cite{Yanik2004}, and could be practically implemented in lithographically designed optomechanical systems both in the microwave \cite{Regal2008} and optical \cite{Eichenfield2009} domain.

{\small $\hphantom{xxx}$\newline\noindent\textbf{Acknowledgements}
 T.J.K. acknowledges financial support by an ERC Starting Grant (SiMP), MINOS, a Marie Curie Excellence Grant, the NCCR of Quantum Photonics of the SNF, and appreciated continued support by MPQ. \newline}

{\small \noindent\textbf{Author contributions}
AS contributed the theoretical idea and suggested the measurements. RR, SW and SD constructed the cryogenic setup and performed the measurements. EG fabricated the employed samples. OA contributed at an early stage of the experiment. AS, SW, SD and TJK analyzed the data and wrote the manuscript. }

\bibliographystyle{unsrt}
\bibliography{/Users/aschlies/Documents/Literature/microCavities}

\widetext

{\begin{figure}[tb]
\begin{center}
{\includegraphics[width=.4\linewidth]{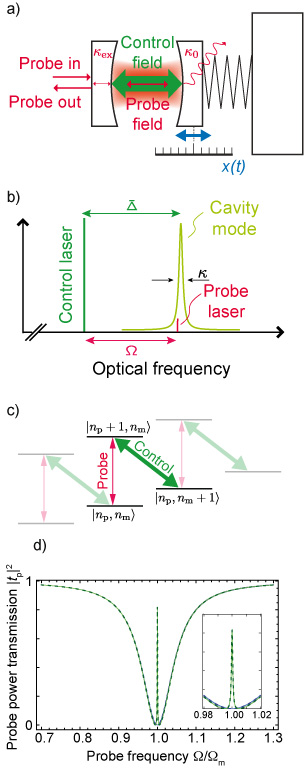}  }
\end{center}
\caption{
Optomechanically induced transparency.
(a) A  generic optomechanical system consists of an optical cavity with a movable boundary, illustrated here as a Fabry-Perot-type resonator in which one mirror acts like a mass-on-a-spring movable along $x$. 
The cavity has an intrinsic photon loss rate $\kappa_0$ and is coupled to an external propagating mode at the rate $\kappa_\mathrm{ex}$.
Through the external mode, the resonator is populated with a control field (only intracavity field is shown).
The response of this driven optomechanical system is probed by a weak probe field sent towards the cavity, the transmission of which  (i.e. the returned field ``Probe out'') is analyzed here.
(b) The frequency of the control field is detuned by $\bar \Delta$ from the cavity resonance frequency, where a detuning close to the lower mechanical sideband, $\bar \Delta\approx -\Omega_\mathrm{m}$, is chosen. The probe laser's frequency is offset by the tunable radio frequency $\Omega$ from the control laser. The dynamics of interest occur when the probe laser is tuned over the optical resonance of the cavity, which has a linewidth of $\kappa= \kappa_\mathrm{0}+ \kappa_\mathrm{ex}$.
(c) Level scheme of the optomechanical system. The control field is tuned close to red-sideband transitions, in which a mechanical excitation quantum is annihilated (mechanical occupation  $n_{\mathrm{m}}\rightarrow n_{\mathrm{m}}-1$) when a photon is added to the cavity (optical occupation $n_\mathrm{p}\rightarrow n_\mathrm{p}+1$), therefore coupling the corresponding energy eigenstates. The probe field probes transitions in which the mechanical oscillator occupation is unchanged. 
(d) Transmission of the probe laser power through the optomechanical system in the case of a critically coupled cavity $\kappa_0=\kappa_\mathrm{ex}$ as a function of normalized probe frequency offset, when the control field is off (blue lines) and on (green lines). Dashed and full lines correspond to the models based on the full (eq.\ (1)) and approximative (eq.\ (5)) calculations, respectively.
}%
\label{f:1}%
\end{figure}}

\begin{figure}[tb]
{\includegraphics[width=\linewidth]{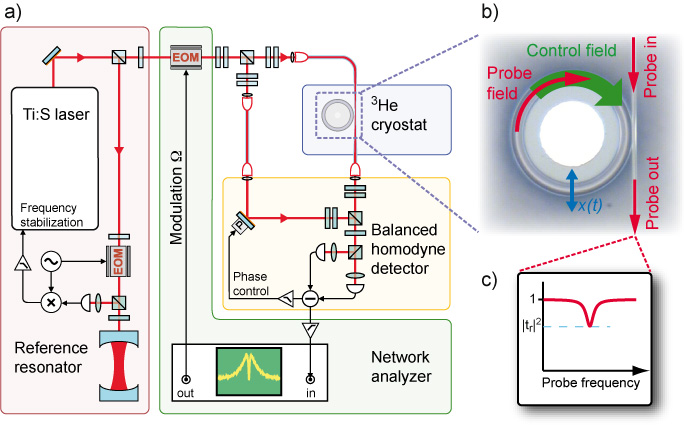}  }
\caption{
Experimental setup. a) An optomechanical system consisting of a toroid microresonator held at cryogenic temperatures in a Helium-3 buffer gas cryostat. The control and probe fields are derived from a single Ti:sapphire laser, which is stabilized to an external reference resonator using the Pound-Drever-Hall technique. While the laser carrier is used as control beam, the probe beam is created by a phase modulator driven at the radio frequency $\Omega$. This optical input is split into two arms, one of which is sent to a tapered fiber in the cryostat, which allows optical coupling to the whispering gallery mode of the toroidal resonator as shown in micrograph with a  $\sim60\,\mathrm{\mu m}$-diameter toroid. The other arm serves as the local oscillator in a balanced homodyne receiver used to analyze the light returned from the optomechanical system. While the receiver's DC component is used to lock the phase of the local oscillator, the AC component is analyzed using a network analyzer. c) In absence of the control beam, the transmission of the probe is a simple Lorentzian. Under the chosen waveguid-toroid coupling conditions, there is a non-zero probe power transmission $|t_\mathrm{r}|^2$ at resonance.}
\end{figure}

{\begin{figure}[tb]
\begin{center}
\includegraphics[width=\linewidth]{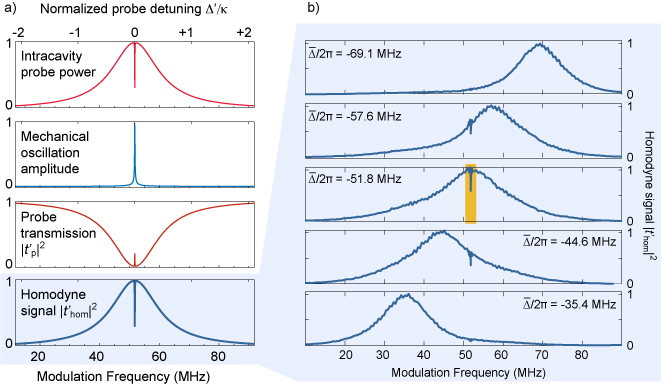}
\end{center}
\caption{
Observation of OMIT. a) Theoretically expected intracavity probe power $|A^-|^2$,
oscillation amplitude $X$, probe power transmission $|t_\mathrm{p}|^2$ and the homodyne signal $|t_\mathrm{hom}|^2$ as a function of the modulation frequency $\Omega/2\pi$ (top to bottom panels). The first two panels have additionally been normalized to unity. When the two-photon resonance condition is met, the mechanical oscillator is excited $\Delta'=0$, giving rise to destructive interference of excitation pathways for an intracavity probe field. The probe transmission therefore exhibits an inverted dip, which can be easily identified in the homodyne signal. b) Experimentally observed normalized homodyne traces when the probe frequency is scanned by sweeping the phase modulator frequency $\Omega$ for different values of control beam detuning $\bar \Delta$. While the center of the response of the bare optical cavity shifts correspondingly, the sharp dip characteristic of OMIT occurs always for $\Delta'=0$. The power of the control beam sent to the cavity is $0.5\ \mathrm{mW}$ in these measurements and the He-3 buffer gas has a pressure of $155\, \mathrm{mbar}$  at a temperature of  $3.8\, \mathrm{K}$. The middle panel shows the operating conditions where the control beam is tuned to the lower motional sideband $\bar \Delta\approx-\Omega_\mathrm{m}=-2\pi\cdot 51.8\,\mathrm{MHz}$. The region around the central dip (orange background) is studied in more detail in a dedicated experimental series (cf.\ figure 4).}%
\end{figure}}

{\begin{figure}[tb]
{\includegraphics[width=\linewidth]{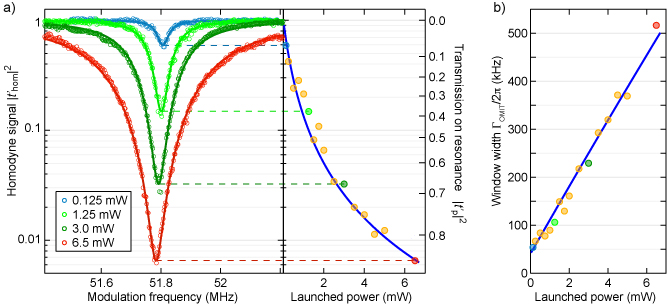}
 }
\caption{
Controlling optomechanically induced transparency. 
a) Experimental normalized homodyne traces in the presence of a control beam (circles) for four different powers in the control beam from $0.125\, \mathrm{mW}$ up to $6.5\,\mathrm{mW} $, and Lorentzian models. The minimum homodyne signal (measured at $\Delta'=0$) directly indicates the maximum probe power transmission achieved in this case. These values are given in the right panel for a larger set of probe scans, together with the theoretical model developed in this work. See text for more information.
b) Width of the transparency window extracted from the same set of probe scans. Good agreement with the theoretical prediction is found over the entire power range. }%
\label{setup}%
\end{figure}}

\clearpage


\setcounter{figure}{0}%
\setcounter{equation}{0}%
\setcounter{section}{0}
\renewcommand \theequation {S\arabic{equation}}%
\renewcommand \thefigure {S\arabic{figure}}
\renewcommand \thesection {S \arabic{section}}

\begin{center}
\large{
\textbf{Appendix - Optomechanically induced transparency}}
\end{center}
\vspace{.2in}

\section{Derivation of OMIT}
In the following, we derive the expressions describing the optomechanical equivalent of
Electromagnetically Induced Transparency (EIT), as discussed in
\cite{SISchliesser2009,SISchliesser2010}, as well as more recent
analysis \cite{SIAgarwal2010}. The starting point of the
following analysis is the Hamiltonian formulation of a generic
optomechanical system put forward by Law \cite{SILaw1995}.

\subsection{Hamiltonian}
If the free spectral range of the cavity is much larger than the
mechanical oscillation frequency, such that only one optical mode is
coupled to the mechanical mode, the optomechanical Hamiltonian can be written as: 
\begin{align}
\hat H&=\hat H_\mathrm{mech}+\hat H_\mathrm{opt}+\hat H_\mathrm{int}+\hat H_%
\mathrm{drive}  \label{e:h} \\
\hat H_\mathrm{mech} &= \frac{\hat p ^2}{2 m_\text{eff}}+ \frac{1}{2} m_%
\text{eff} \Omega_\mathrm{m} ^2 \hat x ^2 \\
\hat H_\mathrm{opt}&=\hbar \omega_\mathrm{c}  \left(\hat a^\dagger  \hat a +%
\frac{1}{2}\right) \\
\hat H_\mathrm{int}&=\hbar g_0  \hat x \,\hat a^\dagger  \hat a  \\
\hat H_\mathrm{drive}&= {i \hbar} {\ \sqrt{\etac  \kappa}}
\left(s_\mathrm{in}(t)  \hat a^\dagger -s_%
\mathrm{in} ^*(t) \hat a \right),
\end{align}
where $\hat x$ and $\hat p$ are the position and momentum operators of
the mechanical degree of freedom having effective mass
$m_{\mathrm{eff}}$ and angular frequency $\Omega_m$, and $s_\mathrm{in}(t)$ is the
drive amplitude normalized to a photon flux at the input of the
cavity.
$\hat a$ and $\hat a^\dagger$ are the annihilation and creation operators of the cavity mode. 
We have furthermore used the coupling parameter $\etac \equiv
{\kappa_\mathrm{ex}}/{\kappa_0 + \kappa_\mathrm{ex}}$, where
$\kappa_0$ denotes the intrinsec loss rate and $\kappa_\mathrm{ex}$
the external loss rate (i.e. wave guide
coupling). Experimentally, the parameter $\etac$ can be continuously adjusted by
tuning the taper-resonator gap \cite{SICai2000,SISpillane2003}.

We will solve this problem for a driving field $s_\mathrm{in}(t)=\left(\bar
s_\mathrm{in}  + \delta s_\mathrm{in}(t)\right)e^{-i \omega_\mathrm{l} t}$,
where $\omega_{\mathrm{l}}$ is the driving laser frequency, and we deliberately identify
$\bar s_\mathrm{in}=s_\mathrm{l}$. We will then first derive the linearized Langevin equations 
\cite{SIFabre1994} for
a generic perturbation term $\delta s_\mathrm{in}(t)$ before identifying it with the probe
field $\delta s_\mathrm{in}(t)=s_\mathrm{p} e^{-i (\omega_\mathrm{p} -
 \omega_\mathrm{l} ) t}$.
\subsection{Langevin equations}
In a frame rotating at $\omega_\mathrm{l} $ with
$\Delta=\omega_\mathrm{l}-\omega_\mathrm{c}$, neglecting quantum and
thermal noise we obtain:
\begin{align}
\frac{d}{dt}\hat a (t)&=\left(+i \Delta-\frac{\kappa}{2}\right) \hat a (t)-
i g_0  \hat x (t) \hat a (t) + \sqrt{\etac \kappa} s_\mathrm{in}(t)  \label{e:aQLE}\\
\frac{d}{dt}\hat x (t)&=\frac{\hat p (t)}{m_\text{eff}}  \label{e:pQLE} \\
\frac{d}{dt}\hat p (t)&=-m_\text{eff} \Omega_\mathrm{m} ^2 \hat x (t)- \hbar
g_0  \hat a^\dagger (t) \hat a (t)-{\Gamma_\mathrm{m} } \hat p (t), \label{e:xQLE}
\end{align}
where the decay rates for the optical ($\kappa$) and mechanical
oscillators ($\Gamma_m$) have been introduced classically. We first
denote $\bar a$ and $\bar x$ the intracavity
field and mechanical displacement for the static solution, in which all time derivatives vanish and $s_\mathrm{p}\rightarrow 0$.
 From \eqref{e:aQLE}\eqref{e:xQLE}, it follows
immediately that $\bar a$ and $\bar x$ must fulfill the self
consistent equations:
\begin{align}
\bar a &= \frac{\sqrt{\etac \kappa}}{-i (\Delta - g_0 \bar x) + \kappa/2}  \bar
s_\mathrm{in} \\
\bar x &= \frac{{\bar a}^2}{m_\mathrm{eff} \Omega_m^2},
\end{align}
where we have assumed $a$ to be real and positive.
This system can give rise to bistability for sufficiently
strong control fields \cite{SIDorsel1983}\cite{SIFabre1994}. However, for weak and
detuned control fields, only one solution exists and ${|\bar a|}^2
\propto \etac {\bar{s_\mathrm{in}}}^2$. We then linearize
the problem for $\delta s_\mathrm{in} \ll |\bar s_\mathrm{in}|$,
plugging the ansatz $\hat a (t)=\bar a +\delta \hat a (t)$ and $\hat x
(t)=\bar x +\delta \hat x (t)$ into equations
\eqref{e:aQLE}\eqref{e:pQLE}\eqref{e:xQLE} and retain only first order terms in the
small quantities $\delta \hat a $, $\delta \hat a^{\dagger} $ and
$\delta \hat x $.  We then obtain
\begin{align}
\label{e:aQLElin}
\frac{d}{dt}\, \delta \hat a (t) &=\left(+i \bar \Delta  - \frac{\kappa}{2}%
\right)\delta \hat a (t)-i g_0 \bar a  \delta \hat x (t)+
\sqrt{\etac \kappa} \delta s_\mathrm{in}(t)\\
\frac{d^2}{dt^2}\, \delta \hat x (t)&+ \Gamma_\mathrm{m}  \frac{d}{dt} \,
\delta \hat x (t)+\Omega_\mathrm{m} ^2 \delta \hat x (t)=-\frac{\hbar g_0}{m_%
\text{eff}}\bar a \left( \delta \hat a (t)+\delta \hat a^{\dagger}
(t)\right),\label{e:xQLElin}
\end{align}
where we used the Hermitian property $\delta \hat x (t)=\delta \hat
x^{\dagger} (t)$ in equation \eqref{e:aQLElin}, and introduced the corrected detuning $\bar \Delta =
\Delta - g_0 \bar x$. 
Since the drives are weak, but classical coherent fields, we will
identify all operators with their expectation
values $y(t) \equiv \langle \hat y(t) \rangle$.

\subsection{Solution}

\subsubsection{General solution}
We now have to solve the equations of the expectation values for the drive (in the rotating frame) $\delta
s_\mathrm{in}(t)=s_\mathrm{p} e^{-i (\omega_\mathrm{p} -
  \omega_\mathrm{l})t}$. 
For a given $\Omega=\omega_\mathrm{p}-\omega_\mathrm{l}$ we use the ansatz 
\begin{align}
\delta a(t)&=A^{-}\, e^{-i \Omega t}+A^+\,e^{+i \Omega t}\\
\delta a^*(t)&=(A^+)^*\, e^{-i \Omega t}+(A^-)^*\,e^{+i \Omega t}\\ 
\delta x(t)&=X \, e^{-i \Omega t}+X^*\,e^{+i \Omega t}.
\end{align}
If sorted by rotation terms, this yields six equations.
However, the probe field's transmission at frequency $\omega_l +
\Omega$ only depends on $A^-$. In this sense, the three
equations of interest are:
\begin{align}  
\left(-i (\bar \Delta +\Omega ) +\kappa/2\right)A^-
&=-ig_0 \bar a  X+
\sqrt{\etac \kappa} s_p  \label{e:eomx1}
\\
\left(+i (\bar \Delta -\Omega ) +\kappa/2\right)(A^+)^*
 &=+ig_0 \bar a  X   \label{e:eomx2}
\\
m_\text{eff}\left(\Omega_\mathrm{m} ^2-\Omega ^2-i \Gamma_\mathrm{m}  \Omega
\right) X &= -\hbar g_0  \bar a  \left(
A^- + (A^+)^*\right). \label{e:eomx3}
\end{align}
The solution of interest is
\begin{align}
A^-&=\frac{1+i f(\Og)}
        {-i(\bD+\Og)+\kappa/2+2\bD f(\Og)} \sqrt{\etac \kappa} s_p, \intertext{with}
  f(\Og)&=\hbar \dwdx^2 \ba^2\,\frac{\chi(\Og)}{i(\bD-\Og)+\kappa/2}
  \intertext{and the mechanical susceptibility}
\chi(\Omega)&=\frac{1}{m_\mathrm{eff}}\frac{1}{\Omega_\mathrm{m}^2-\Omega^2-i\Omega \Gamma_\mathrm{m}}.
\end{align}

\subsubsection{Spectrum of the transmitted light}
Using the input-output relation \cite{SIGardiner2004}, one obtains:
\begin{align}
s_\mathrm{out}(t)
	&=s_\mathrm{in}(t)-\sqrt{\etac \kappa}\, a(t)\\
	&=
		(s_\mathrm{c}-\sqrt{\etac \kappa} \bar a) e^{-i \omega_c t}+
		(s_\mathrm{p}-\sqrt{\etac \kappa} A^-) e^{-i (\omega_c+\Omega) t}
		-\sqrt{\etac \kappa} A^+ e^{-i (\omega_c-\Omega) t}.
\end{align}
The transmission of the probe beam, defined by the ratio of the output
and input field amplitudes at the probe frequency is then given by:
\begin{align}
\label{e:transmission}
t_p &= 1-\sqrt{\etac \kappa} A^-\\
    &= 1- \frac{1+i f(\Omega)}{-i(\bar \Delta +
  \Omega) + \kappa/2 + 2 \bar \Delta f(\Omega)}\etac \kappa.
\end{align}
\subsubsection{Resolved-sideband limit}
In the resolved sideband regime \cite{SIWilson-Rae2007} ($\kappa \ll
\Omega_m$), the lower sideband, far off-resonance,  can be neglected:
\[
A^+ \approx 0.
\]
In addition, we can linearize the mechanical susceptibility
for small values of the parameter $\Delta^{\prime} = \Omega-\Omega_m$:
\[
m_{\mathrm{eff}}(\Omega_m^2-\Omega^2-i \Gamma_m \Omega) \approx \Omega_m (2
\Delta^{\prime} - i \Gamma_m).
\]
The system \eqref{e:eomx1}\eqref{e:eomx2}\eqref{e:eomx3} then simplifies to:
\begin{align}
\left(-i (\bar \Delta + \Omega_m + \Delta^{\prime}) +\kappa/2\right)A^-
&=-ig_0 \bar a  X+
\sqrt{\etac \kappa} s_p  \label{e:eomxsb1}
\\
\Omega_m (2\Delta^{\prime} - i \Gamma_m) X &= -\hbar g_0  \bar a
A^-. \label{e:eomxsb2}
\end{align}
The solution for the intracavity field is:
\begin{equation}
\label{e:rsb1}
A^- = \frac{\sqrt{\etac \kappa} s_p}{-i (\bar \Delta + \Omega_m + \Delta^{\prime}) + \kappa
  /2 + \frac{\Omega_c^2/4}{ ( -i \Delta^{\prime} +\Gamma_m/2)}},
\end{equation}
where we have introduced the coupling between the mechanical and
optical resonators:
\[
\Omega_c = 2 g_0 \bar a x_\mathrm{zpf},
\]
with
\[
 x_\mathrm{zpf} = \sqrt{\frac{\hbar}{2 m_\mathrm{eff} \Omega_m}},
\]
the zero point fluctuations amplitude of the mechanical oscillator. 
This formula becomes
\begin{equation}
\label{e:rsb2}
A^- = \frac{\sqrt{\etac \kappa} s_p}{-i \Delta^{\prime} + \kappa
  /2 + \frac{\Omega_c^2/4}{ ( -i \Delta^{\prime} + \Gamma_m/2)}}
\end{equation}
for a control laser tuned to the lower motional sideband ($\bar \Delta = - \Omega_m$).
We will now briefly review the formalism used to describe EIT in the context of atomic
physics to emphasize the analogy between the two phenomena.

\section{Atomic EIT}
For atomic EIT, we essentially revisit the well-known derivation developed in
\cite{SIMilonni2005}, which will assist in identifying the close
resemblance between OMIT and atomic EIT. We consider a $\Lambda$-system, consisting of a common upper state $|3\rangle$ and two (long-lived) ground states $|1\rangle$ and $|2\rangle$.
In a semiclassical treatment \cite{SIScully1997}, the relevant Hamiltonian
is given by:
\begin{equation}
  \hat H_\mathrm{}=\sum_j \hbar \omega_j \hat \sigma_{jj}
  -\frac{\hbar}{2}\mu_{23}(\hat \sigma_{23}+\hat\sigma_{32}) E(t) 
  -\frac{\hbar}{2}\mu_{13}(\hat \sigma_{13}+\hat\sigma_{31}) E(t),
\end{equation}
where $\hat\sigma_{ij}= |i\rangle\langle j|$ are the atomic projection
operators and $i,j\in\{1,2,3\}$ label the three involved
levels. $\omega_{ij}$ and $\mu_{ij}$ are the frequency and dipole moment along the electric field's
direction for the $i \rightarrow j$ transition.
The (classical) field contains the two (coupling and probe) components
\begin{equation}
 E(t)=\frac{1}{2}\mathcal{E}_\mathrm{c}\left(e^{-i \omega_\mathrm{c} t}+e^{+i \omega_\mathrm{c} t}\right)
       +\frac{1}{2}\mathcal{E}_\mathrm{p}\left(e^{-i \omega_\mathrm{p} t}+e^{+i \omega_\mathrm{p} t}\right),
\end{equation}
where $ \omega_\mathrm{c}$ is tuned close to $ \omega_{32}$ and $ \omega_\mathrm{p}$ close to $ \omega_\mathrm{31}$.
The usual Heisenberg equations of motion for the operators $\hat\sigma_{ij}$ can then be derived using 
$i\hbar \frac{d {{\hat \sigma}}_{ij}}{dt}=[{\hat \sigma}_{ij},\hat H]$. 
Retaining only near-resonant terms, the equations of motion can be written as 
\begin{align}
{\dot {\hat \sigma}}_{12}&=-i\omega_{21} \hat \sigma_{12}
	+\frac{i}{2\hbar}\mu_{23}\mathcal{E}_\mathrm{c}\hat\sigma_{13} e^{+i\omega_\mathrm{c} t}
	-\frac{i}{2\hbar}\mu_{13}\mathcal{E}_\mathrm{p}\hat\sigma_{32} e^{-i\omega_\mathrm{p} t}\\
{\dot {\hat \sigma}}_{23}&=-i\omega_{32} \hat \sigma_{23}
	+\frac{i}{2\hbar}\mu_{23}\mathcal{E}_\mathrm{c}(\hat\sigma_{22}-\hat\sigma_{33}) e^{-i\omega_\mathrm{c} t}
	+\frac{i}{2\hbar}\mu_{13}\mathcal{E}_\mathrm{p}\hat\sigma_{21} e^{-i\omega_\mathrm{p} t}\\
{\dot {\hat \sigma}}_{13}&=-i\omega_{31} \hat \sigma_{13}
	+\frac{i}{2\hbar}\mu_{23}\mathcal{E}_\mathrm{c}\hat\sigma_{12} e^{-i\omega_\mathrm{c} t}
	+\frac{i}{2\hbar}\mu_{13}\mathcal{E}_\mathrm{p}(\hat\sigma_{11}-\hat\sigma_{33}) e^{-i\omega_\mathrm{p} t}.
\end{align}
We emphasize that the rotating wave approximation (neglecting all
non-resonant contributions) is analogous to the resolved sideband approximation presented
in the context of OMIT.
For a sufficiently weak probe field, the expectation values $\sigma_{ij}=\langle \hat \sigma_{ij}\rangle$ can further be approximated to
obey $\sigma_{11}\approx 1$ and $\sigma_{22}\approx \sigma_{33}\approx \sigma_{23}\approx \sigma_{32}\approx 0$
at all times, while the remaining expectation values obey
\begin{align}
{\dot { \sigma}}_{12}&=-i(\omega_{21}-i\gamma_{12}/2)  \sigma_{12}
	+\frac{i}{2\hbar}\mu_{23}\mathcal{E}_\mathrm{c}\sigma_{13} e^{+i\omega_\mathrm{c} t}\\
{\dot { \sigma}}_{13}&=-i(\omega_{31}-i\gamma_{13}/2)  \sigma_{13}
	+\frac{i}{2\hbar}\mu_{23}\mathcal{E}_\mathrm{c}\sigma_{12} e^{-i\omega_\mathrm{c} t}
	+\frac{i}{2\hbar}\mu_{13}\mathcal{E}_\mathrm{p} e^{-i\omega_\mathrm{p} t},
\end{align}
where damping rates $\gamma_{12}$ and $\gamma_{13}$ were introduced classically.
Changing to a rotating frame $ \sigma_{12}=S_{12}e^{-i\Omega t}$,
$ \sigma_{13}=S_{13}e^{-i(\omega_\mathrm{c}+\Omega)t}$ and
$\mathcal{E}_\mathrm{p} e^{-i\omega_\mathrm{p} t}=
\mathcal{E}_\mathrm{p}e^{-i(\omega_\mathrm{c}+\Omega)t}$ with $\omega_\mathrm{p}=\omega_\mathrm{c}+\Omega$,
we obtain in the steady state
\begin{align}
(-i(\Omega-\omega_{21})+\gamma_{12}/2)S_{12}&=+\frac{i}{2\hbar}\mu_{23}\mathcal{E}_\mathrm{c}S_{13} \label{e:eit1} \\
(-i(\Omega+\omega_c-\omega_{31})+\gamma_{13}/2)  S_{13}&=
	+\frac{i}{2\hbar}\mu_{23}\mathcal{E}_\mathrm{c}S_{12} 
	+\frac{i}{2\hbar}\mu_{13}\mathcal{E}_\mathrm{p}\label{e:eit2},
\end{align}
which is solved by
\begin{equation}
 \label{e:eit3}
  S_{13}=\frac{i \mu_{13} \mathcal{E}_\mathrm{p}/2 \hbar}{-i
    (\Delta^{\prime} + \omega_c - \omega_{32})+\gamma_{13}/2+\frac{\Omega_\mathrm{c}^2/4}{-i \Delta^{\prime}+\gamma_{12}/2}},
\end{equation}
where we now abbreviate $\Delta^{\prime}=\Omega-\omega_{21}=\omega_\mathrm{p}-\omega_{31}$.
We note as an aside that an equivalent calculation can be made for the
atomic coherences $\rho_{12}$ and $\rho_{13}$, yielding essentially
the same result \cite{SIScully1997}. This result simplifies for a control
field on resonance ($\omega_c = \omega_{32}$):
\begin{align}
\label{e:eit3res}
  S_{13}=\frac{i \mu_{13} \mathcal{E}_\mathrm{p}/2 \hbar}{-i
    \Delta^{\prime}+\gamma_{13}/2+\frac{\Omega_\mathrm{c}^2/4}{-i \Delta^{\prime}+\gamma_{12}/2}}.
\end{align}
The induced dipole moment along the electric field's direction is given by $p=\mu_{13}(\sigma_{13}+\sigma_{31})$
so that the polarizibility $\alpha$ of the medium at the probe frequency in the presence of the coupling beam can be directly given by
\begin{equation}
\label{e:omitres}
  \alpha=\frac{\mu_{13} S_{13}}{\mathcal{E}_\mathrm{p}/2}=
 		 \frac{i \mu_{13}^2 / \hbar}{(-i\Delta^{\prime}+\gamma_{13}/2)+\frac{\Omega_\mathrm{c}^2/4} {(-i\Delta^{\prime}+\gamma_{12}/2)}}.
\end{equation}
Evidently one can identify a formal correspondence between the
physical entities involved in EIT in atomic physics and OMIT in
optomechanical systems. Equations \eqref{e:eit1}\eqref{e:eit2}\eqref{e:eit3}\eqref{e:eit3res} are
perfectly equivalent to \eqref{e:eomxsb1}\eqref{e:eomxsb2}\eqref{e:rsb1}\eqref{e:rsb2} by applying
the identifications listed in the following table.

\begin{table}[htdp]
\caption{Comparison of physical entities relevant for EIT and OMIT.}
\begin{center}
\begin{tabular}{m{7.5cm} || m{6.5cm}}
  EIT  & OMIT\\
  \hline
   projection operator $\sigma_{13}$ (coherence $\rho_{13}$)	& intracavity field amplitude $A^-$ \\
   projection operator $\sigma_{12}$ (coherence $\rho_{12}$)	& mechanical displacement amplitude $X$ \\
   energy difference between ground states $ \hbar \omega_{21}$  	& phonon energy $\hbar \Omega_\mathrm{m}$\\
   Rabi frequency $\mu_{23}\mathcal{E}_\mathrm{c}/\hbar$		& optomechanical coupling rate $2 g_0 \bar a x_\mathrm{zpf}$
\end{tabular}
\end{center}
\label{default}
\end{table}%

\section{Simplified expressions in the weak coupling case}
In addition to the resolved sideband approximation, we will
consider the case where the optomechanical coupling is weak compared
to the optical losses ($\Omega_c,\Gamma_m \ll \kappa$). We also assume
that the control laser is tuned on the lower
sideband ($\bar \Delta = -\Omega_m$). Then, the EIT feature is
very well described by a Lorentzian transmission window in the
optical transmission spectrum. This can be seen by applying the
simplification $-i\Delta^{\prime}+\kappa/2 \approx \kappa/2$ in equation
\eqref{e:rsb2}:
\begin{equation}
A^- \approx \frac{4 \sqrt{\etac \kappa} (-i \Delta^{\prime} + \Gamma_m/2)}{2 \kappa
  (\Gamma_m/2 - i \Delta^{\prime}) + \Omega_c^2} s_p.
\end{equation}
Plugging the corresponding value of $A^-$ in
\eqref{e:transmission}, one obtains:
\begin{equation}
\label{e:diplorentz}
t_p = 1 - 2 \etac + \frac{2 \etac \Omega_c^2}{\Omega_c^2+\Gamma_m \kappa-2 i \Delta^{\prime} \kappa}.
\end{equation}
In order to isolate the interesting physics of OMIT from the well-understood waveguide-cavity coupling effects, we introduce the normalized transmission:
\begin{align}
t_p^{\prime} = \frac{t_p-t_r}{1-t_r},
\end{align}
where $t_r$ is the residual on resonance transmission in the absence
of a coupling laser:
\begin{align}
t_r &= t_p(\Delta^{\prime} = 0,\Omega_c = 0)\\
    &=1-2 \etac. \label{e:residual}
\end{align}
The normalized transmission is then independant of $\etac$:
\begin{align}
\label{e:diplorentznormalized}
t_p^{\prime} = \frac{\Omega_c^2}{\Omega_c^2+\Gamma_m \kappa-2 i \Delta^{\prime} \kappa}.
\end{align}
This corresponds to the transmission in the case of critical coupling
$\etac = 1/2$. The optomechanically induced transparency window is
hence given by:
\begin{align}
|t_p^{\prime}|^2 =
\frac{\Omega_c^4/\kappa^2}{\left(\Omega_c^2/\kappa+\Gamma_m\right)^2 +  {\left(2 \Delta^{\prime}\right)}^2}.
\end{align}
A Lorentzian of width
\begin{align}
\Gamma_\mathrm{OMIT} &= \Gamma_m + \Omega_c^2/\kappa\\
\end{align}
and peak value
\begin{align}
|t_p^{\prime}(\Delta^{\prime} = 0)|^2 &= \left(\frac{\Omega_c^2/\kappa}{ \Gamma_m + \Omega_c^2/\kappa}\right)^2.
\end{align}
These two quantities can be expressed very simply by introducing the
cooperativity of the coupled systems $C = \Omega_c^2/(\Gamma_m
\kappa)$ \cite{SIKimble1994}:
\begin{align}
\Gamma_\mathrm{OMIT} &= \Gamma_m(1 + C)\\
|t_p^{\prime}(\Delta^{\prime} = 0)|^2 &= \left(\frac{C}{ 1 + C}\right)^2.
\end{align}

\section{Measurement using the phase modulation scheme}
For technical reasons, the optical response was probed
using a frequency modulation technique: the coupling laser is phase
modulated using an EOM at frequency $\Omega$, hence creating two
sidebands at $\omega_l + \Omega$ and $\omega_l - \Omega$. 
In the resolved sideband regime, only the upper sideband,
close to resonance interacts with the cavity, acting as a weak probe beam. The lower one and carrier are
transmitted unchanged through the tapered fiber. However, one has to
take them into account in order to understand quantitatively the
obtained results. We will show here that the measured signal is linked
to the transmission at the probe frequency through a 
direct relation.

The precise measurement scheme is given in figure 2, and reproduced in figure \ref{f:setupSI}. The incident fields at the homodyne beamsplitter are
a carrier and two sidebands of the local oscillator, and the carrier and two sidebands
of the beam entering the cavity. We note $t_c$, $t_{\mathrm{us}}$
and $t_{\mathrm{ls}}$ the complex transmission coefficient across the
cavity for the carrier, upper sideband and lower sideband respectively.
The phase of the local oscillator $\Phi$ is actively adjusted so that it matches the phase of the control beam emerging from the cavity.

{\small \begin{figure}[tb]
\begin{center}
{\small \includegraphics[width=.8\linewidth]{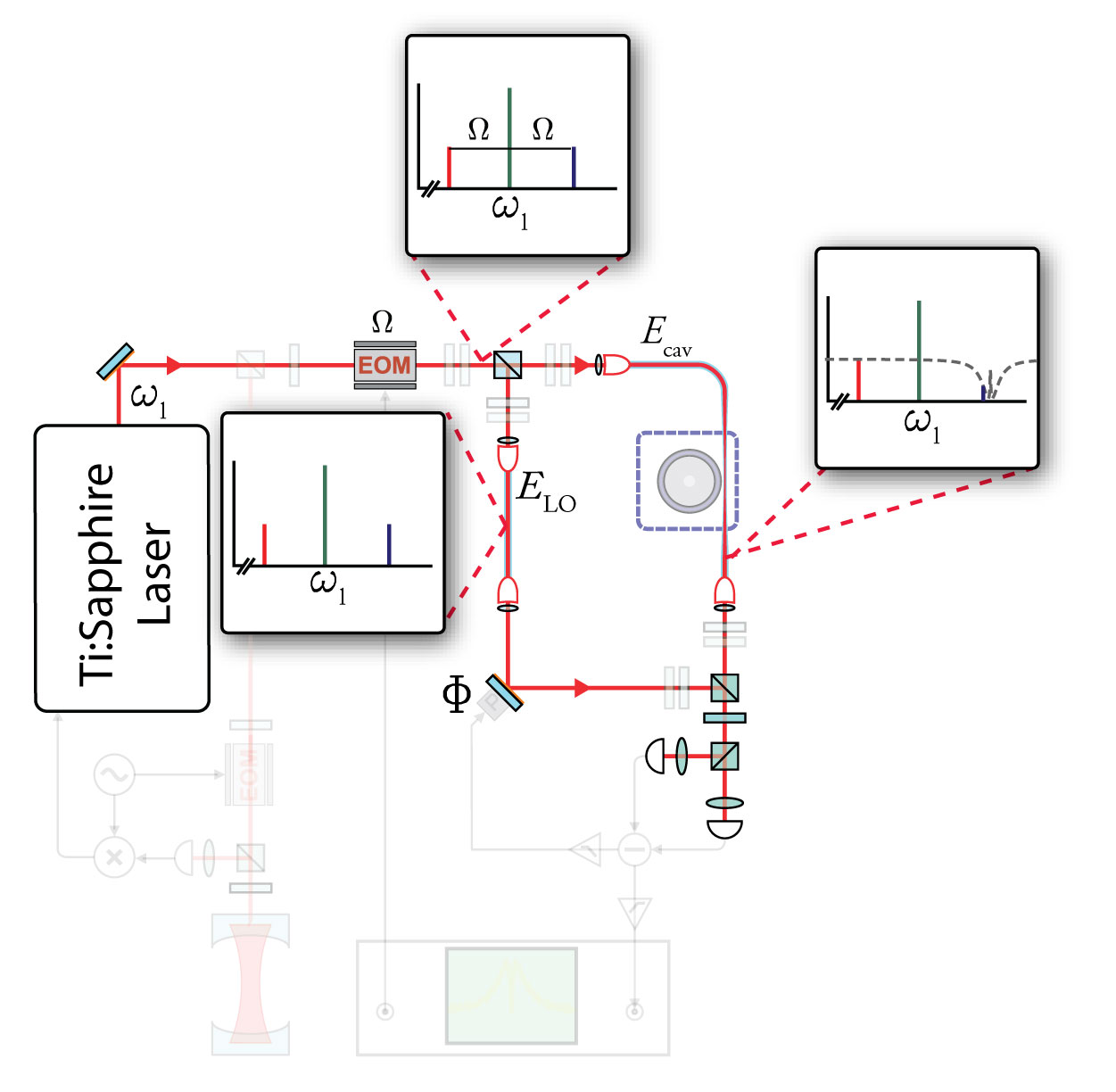}  }
\end{center}
\caption{The optical setup as described in the main manuscript. The
  laser is phase modulated, creating two sidebands at frequency
  $\omega_l \pm \Omega$. The local
  oscillator field is transmitted unchanged whereas the field in the
  signal arm is affected by the cavity transmission. In the RSB
  regime, lower sideband and carrier, off resonant by approximately $2 \Omega_m$ and
  $\Omega_m$ are not affected.}
\label{f:setupSI}
\end{figure}}

At one exit of the beamsplitter the optical power is proportional to
\begin{align}
\left |
   E_{\mathrm{cav}} e^{i \omega_l t} \left (
   t_{\mathrm{c}} +
   i\frac{\beta}{2} e^{+i \Omega t}t_{\mathrm{us}}+
   i\frac{\beta}{2} e^{- \Omega t}t_{\mathrm{ls}}
   \right)+
   i E_\mathrm{LO}e^{i \omega_l t}e^{i \Phi} \left(1 + i\frac{\beta}{2}
   e^{+i \Omega t} + i\frac{\beta}{2}  e^{-i \Omega t}\right)   
\right |^2,
\end{align}
where $\beta$ is the depth of the modulation induced by the EOM, $E_\mathrm{LO}$ and $E_{\mathrm{cav}}$ are the field amplitudes in
the local oscillator and signal arms of the homodyne setup. 
The interesting terms are the modulated cross-terms, they are given by
\begin{align}
\begin{split}2 \,\mathrm{Re}
   &\left[ \left(E_{\mathrm{cav}} e^{i \omega_l t} t_{\mathrm{c}}\right)
   \cdot
   \left(i E_{\mathrm{LO}}e^{i \omega_l t}e^{i \Phi} \left(
   i\frac{\beta}{2}  e^{+i \Omega t}
+  i\frac{\beta}{2}  e^{-i \Omega t}\right)\right)^* \right.\\
&\qquad \qquad+
\left.
\left(
 E_{\mathrm{cav}} e^{i \omega_l t} 
 \left(
     i\frac{\beta}{2}  e^{+i \Omega t}t_{\mathrm{us}}+
     i\frac{\beta}{2}  e^{-i \Omega t}t_{\mathrm{ls}}
   \right)\right)\cdot
   \left(i E_\mathrm{LO} e^{i \omega_l t}e^{i \Phi}\right)^*
\right]\\
&=\beta  E_{\mathrm{cav}}  E_{\mathrm{LO}} \,\mathrm{Re}
\left[
t_{\mathrm{c}}
   \cdot
   \left(- e^{-i \Phi} \left(e^{-i \Omega t}+ e^{+i \Omega t}\right)\right)+
 \left(
   e^{+i \Omega t}t_\mathrm{us}+
   e^{-i \Omega t}t_\mathrm{ls}
   \right)\cdot
   \left(e^{-i \Phi}\right)
\right]\\
&=\beta E_\mathrm{cav} E_\mathrm{LO} \,\mathrm{Re}
\left[
 \left(e^{-i \Phi}\right)
 \left(
 -t_\mathrm{c}2 \cos(\Omega t)+
 \left(
   e^{+i \Omega t}t_\mathrm{us}+
   e^{-i \Omega t}t_\mathrm{ls}
   \right)\right)
\right]
\end{split}
\end{align}
Now writing real and imaginary parts of the used functions as
\begin{align}
  e^{-i \Phi}&\equiv\Phi'+i\Phi''\\
  t_\mathrm{c}&\equiv t_c'+i t_c''\\
  t_\mathrm{us}&\equiv t_\mathrm{us}'+i t_\mathrm{us}''\\
  t_\mathrm{ls}&\equiv t_\mathrm{ls}'+i t_\mathrm{ls}''
\end{align}
we get (omitting the prefactor $\beta  E_\mathrm{cav}  E_\mathrm{LO} $)
\begin{align}
  \cos(\Omega t)
  \underbrace{
  \left(
  -2\Phi_\mathrm{r} t_\mathrm{c}'+2\Phi'' t_c''
  +( t_\mathrm{us}'+ t_\mathrm{ls}')\Phi'
  -( t_\mathrm{us}''+ t_\mathrm{ls}'')\Phi''
  \right)}_A
  +
   \sin(\Omega t)
  \underbrace{ \left(
 -( t_\mathrm{us}''-t_\mathrm{ls}'')\Phi'
  -( t_\mathrm{us}'- t_\mathrm{ls}')\Phi''
  \right)}_B \label{e:AB}
\end{align}
$A$ and $B$ represent the in-phase and quadrature response of the
system to the input modulation. In the resolved sideband regime, only
the upper sideband is affected by the cavity. In this case $\Phi'=t_{c}'=
t_\mathrm{ls}'=1$ and $\Phi''= t_\mathrm{c}'' =
t_\mathrm{ls}''=0$. Moreover, the upper sideband, close to resonance,
is probing the OMIT signal $t_\mathrm{us} = t_p$. The quadratures measured by the network analyzer are then:
\begin{align}
A &\approx 1-t_{\mathrm{us}}' =  1-\operatorname{Re}(t_p)\\
B &\approx -t_{\mathrm{us}}''  =   -\operatorname{Im}(t_p). 
\end{align}
The complex amplitude response $\thom= A + i B$ as measured on the
network analyzer is hence given in good approximation by:
\begin{align}
\label{e:therelation}
\thom \approx 1-t_p.
\end{align}
The normalized response $\thom^{\prime}$ is directly related to
the normalized transmission ${t_p}^{\prime}$: 
\begin{align}
 \thom^{\prime} &= \frac{\thom}{1-t_r}\\
              &= 1 - {t_p}^{\prime}.
\end{align}
In particular, if we consider the form \eqref{e:diplorentznormalized} for the
probe beam transmission, the measured signal is then given by:
\begin{equation}
\thom^{\prime} = \frac{\Gamma_m \kappa - 2 i \Delta^{\prime}
  \kappa}{\Omega_c^2+\Gamma_m \kappa - 2 i \Delta^{\prime} \kappa}.
\end{equation} 
We can easily calculate the normalized transmitted power:
\begin{align}
|\thom^{\prime}|^2 & = 1 -  \frac{\Omega_c^2/\kappa \left(\Omega_c^2/\kappa + 2
  \Gamma_m\right)}{\left(\Gamma_m + \Omega_c^2/\kappa\right)^2 +
     (2 \Delta^{\prime})^2} 
     \end{align} 
The measured signal is hence an inverted lorentzian peak with width
$\Gamma_\mathrm{OMIT}$ (same width as $|t_p|^2$). 
The minimum value of the dip $|\thom^{\prime}(\Delta^{\prime} = 0)|^2$ can be linked very
easily to the maximum value of the transmission window
$|t_p^{\prime}(\Delta^{\prime} = 0)|^2$ by remarking that for $\Delta^{\prime} = 0$ the transmission
coefficients $t_p$ and $\thom$ are real. The relation
\eqref{e:therelation} gives then:
\begin{align}
|\thom^{\prime}(\Delta^{\prime} = 0)|^2 = \left(1-\sqrt{|t_p^{\prime}(\Delta^{\prime} = 0)|^2}\right)^2
\end{align}

\section{Group delay}
EIT is the underlying phenomenon allowing for slowing down of light
pulses. Indeed, the sharp transparency window in the medium is
accompanied by a very abrupt change of its refractive index leading to
slow group velocities (see \cite{SIMilonni2005} for a detailed analysis
of the phenomenon). In the case
of a single optically active element like an optomechanical device,
the rapid phase dispersion $\phi(\omega) = \arg(t_p(\omega))$ leads to
a `group delay' $\tau_g$ given by:
\begin{align}
\tau_g = - \frac{d \phi}{d \omega}.
\end{align}
A full calculation based on the expression \eqref{e:rsb2} shows that
the group delay diverges for small values of the
transparency. However, in the regime $C \gtrsim 1$, where the medium is not
completely opaque, a simple calculation based on expression
\eqref{e:diplorentznormalized} is perfectly valid:
\begin{align}
\phi(\Delta^{\prime}) = \arctan \left(\frac{2 \Delta^\prime \kappa}{\Omega_c + \Gamma_m \kappa} \right).
\end{align}

This gives for the middle of the transparency window ($\Delta^\prime =
0$):
\begin{align}
\tau_g(\Delta^{\prime}=0) &= \frac{2 \kappa}{\Omega_c^2 + \Gamma_m \kappa}\\
&=\frac{1}{\Gamma_m} \left(\frac{2}{C + 1}\right) \\
&=\frac{2}{\Gamma_\mathrm{OMIT}}.
\end{align}

\section{Theoretical treatment of a splitted resonance}
Our rinf cavity can support two counterpropagating modes which
are frequency degenerate for symmetry reasons. The propagation
direction of the light in the coupling region therefore determines
which mode is excited. However, as noted in early work on
microspheres \cite{SIWeiss1995}, and in theoretical as well as experimental
work \cite{SIGorodetsky2000,SIkippenberg2002,SIMazzei2007}, due to residual scattering of light at the
surface or in the bulk glass, the counterpropagating mode can also be
significantly populated.

{\small \begin{figure}[tb]
\begin{center}
{\small \includegraphics[width=.8\linewidth]{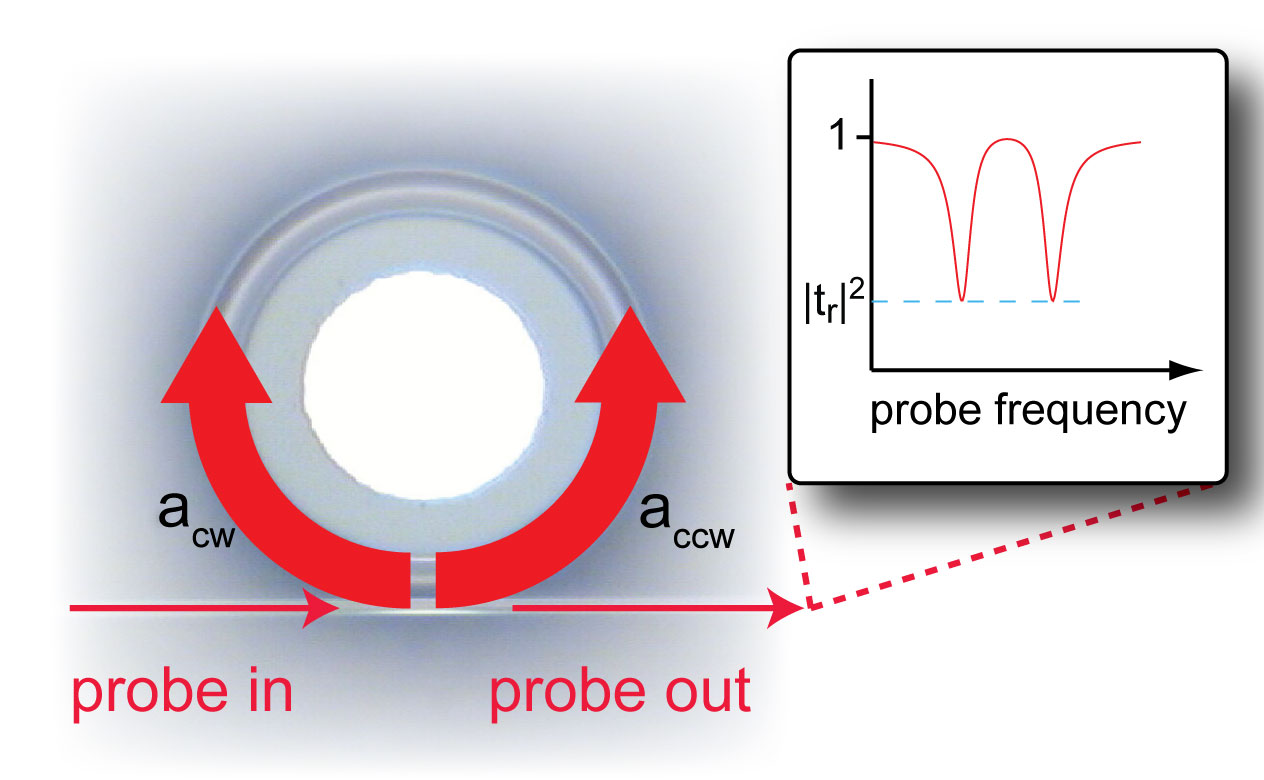}  }
\end{center}
\caption{In a ring cavity, two counterpropagating modes
  $a_\mathrm{cw}$ and $a_\mathrm{ccw}$ coexist. Only $a_\mathrm{ccw}$
  is coupled to the waveguide, however, because of scattering into the
  counterpropagating mode, $a_\mathrm{cw}$ and $a_\mathrm{ccw}$ are
  coupled at a rate $\gamma$. For $\gamma \gg \kappa$, the transmission
spectrum is a splitted resonance.}
\label{f:acwaccw}
\end{figure}}

The essence of the phenomenon can be described by a coupled mode
theory: if the two modes $a_{\mathrm{cw}}$ and $a_{\mathrm{ccw}}$ (see
figure \ref{f:acwaccw}) are
coupled by a coupling rate $\gamma$, the equations of motion become:
\begin{align}
\dot{a}_{\mathrm{ccw}}(t) &= (i (\Delta-g_0 x) - \kappa /2) a_{\mathrm{ccw}}(t)  + i
\frac{\gamma}{2} a_{\mathrm{cw}}(t)+\sqrt{\etac \kappa} s_\mathrm{in}(t)\\
\dot{a}_{\mathrm{cw}}(t) &= (i (\Delta-g_0 x) - \kappa /2)
a_{\mathrm{cw}}(t)  + i
\frac{\gamma}{2} a_{\mathrm{ccw}}(t),
\end{align}
and the radiation pressure force is now described by the equation:
\begin{align}
\frac{d^2}{dt^2} x(t) + \Gamma_m \frac{d}{dt} x(t) + \Omega_m^2 x(t) +
 = -\frac{\hbar g_0}{m_\mathrm{eff}}
(|a_{\mathrm{ccw}}|^2 + |a_{\mathrm{cw}}|^2).
\end{align}
Indeed, because of the symmetry of the radial breathing mode, the
oscillator is not driven by the cross term $2\mathrm{Re}(a_{\mathrm{ccw}}^* a_{\mathrm{cw}})$. 
We can then easily rewrite these equations in terms of the two
stationary modes $a_+ = (a_{\mathrm{ccw}} +
a_{\mathrm{cw}})/\sqrt{2}$ and $a_- =(a_{\mathrm{ccw}} -
a_{\mathrm{cw}})/\sqrt{2}$:
\begin{align}
\dot{a}_+(t) &=  \left[i(\Delta -g_0 x +\frac{\gamma}{2})- \kappa/2 \right] a_+(t)  +
\sqrt{\frac{\etac \kappa}{2}} s_\mathrm{in}(t)\\
\dot{a}_-(t) &= \left[i (\Delta - g_0 x - \frac{\gamma}{2}) - \kappa /2
  \right] a_-(t) + \sqrt{\frac{\etac \kappa}{2}} s_\mathrm{in}(t)\\
\frac{d^2}{dt^2} x(t) + \Gamma_m \frac{d}{dt} x(t) + \Omega_m^2 x(t)  &= -\frac{\hbar g_0}{m_\mathrm{eff}}
(|a_+|^2 + |a_-|^2).
\end{align}
The two stationary modes are the eigenmodes of the evolution and
the degeneracy is lifted by the coupling rate $\gamma$. In the limit
$\gamma \gg \kappa$, the two modes are well
resolved and only one of them ($a_-$) has to be considered since $a_+$
is non resonant and hence not populated. In this limit, the
optomechanical system reads:
\begin{align}
\dot{a}_-(t) &= \left[i (\Delta -g_0 x- \frac{\gamma}{2}) - \kappa /2
  \right] a_-(t) + \sqrt{\frac{\etac \kappa}{2}} s_\mathrm{in}(t)\\
\frac{d^2}{dt^2} x(t) + \Gamma_m \frac{d}{dt} x(t) + \Omega_m^2 x(t)  &= -\frac{\hbar g_0}{m_\mathrm{eff}}
|a_-|^2.
\end{align}
The system is perfectly equivalent to \eqref{e:aQLElin}\eqref{e:xQLElin} by
formally replacing the coupling parameter $\etac$ by $\etac'\equiv\etac/2$. This
reduced `effective coupling parameter` arises from the scattering of half of
the intracavity power to the uncoupled mode $a_{\mathrm{cw}}$.

With our present settings, we measured a residual
transmission of $|t_r|^2 \approx 0.5$, we can hence infer the
effective coupling parameter $\etac'$ by solving
\begin{align}
|t_r|^2 = 1/2 = (1-2 \etac')^2,
\end{align}
leading to
\begin{align}
\etac = \frac{2- \sqrt{2}}{4} \approx 0.15.
\end{align}
The intracavity power is hence smaller than the one calculated in the ``standard'' situation $\etac=1/2$ and $\gamma=0$.
In the calculation of the coupling rate $\Omega_\mathrm{c}$, we took this factor into account; an additional reduction factor of $1.9$ had to be introduced to account for taper losses in this experiment.

\section{Table of symbols}
\centering {
\small
\begin{tabular} {r|p{7.5cm}|l|c}
symbol & meaning & definition \\
\hline %
$\omega_l$   & laser frequency &                                 \\%
$\omega_c$   & cavity resonance frequency                  &     \\%
$\omega_p$   &  probe frequency                             & \\
$\bar \Delta$ & detuning of the control field                  & $\bar \Delta = \Delta - g_0 \bar x
$   \\%
$\kappa$& optical linewidth (FWHM)                         &        \\%
$\kappa_0$ & intrinsic loss rate     &   \\
$\kappa_\mathrm{ex}$ & coupling rate to the waveguide &  \\ 
$\etac$ & coupling parameter   &$\etac=\kappa_\mathrm{ex}/(\kappa_0 + \kappa_{\mathrm{ex}})$      \\%
$ \bar a$  & mean intracavity mode amplitude    &              \\%
$ \bar s_\mathrm{in}$  & mean drive amplitude    &              \\%
%
$\dwdx$   & optomechanical coupling               &      $ d\omega'_\mathrm{c}/dx$      \\%
$\Omega_m$   & mechanical resonance frequency  &                 \\
$\Gamma_m$   & mechanical damping rate         &                 \\
$m_{\mathrm{eff}}$ & effective mass       &                       \\
$\bar x$   & equilibrium displacement  &                       \\%
$\Delta^{\prime}$&detuning of the probe from the center of the OMIT feature &$\Delta^{\prime} = \omega_p -\omega_l - \Omega_m$\\
$x_{\mathrm{zpf}}$ & zero-point fluctuations  & $x_\mathrm{zpf} = \sqrt{\hbar/\left(2 m_\mathrm{eff} \omega_m\right)}$                       \\
$\Omega_c$    & optomechanical coupling rate      & $\Omega_c = 2 g_0
\bar a x_{\mathrm{zpf}}$\\
$C$   & cooperativity      &  $C = \Omega_c^2/(\Gamma_m \kappa)$  \\
$\chi(\Omega)$ & mechanical susceptibility &  $\chi(\Omega)=(m_{\mathrm{eff}}(\Omega_\mathrm{m}^2-\Omega^2-i \Gamma_\mathrm{m} \Omega))^{-1}$ \\%
$t_p$ & complex amplitude transmission at probe frequency & \\
$\thom$ & complex transmission signal measured in the homodyne receiver &\\
$\beta$ & Modulation depth&\\
$t_{\mathrm{us}}$ & complex transmission of the upper sideband&\\
$t_{\mathrm{c}}$ & complex transmission of the carrier&\\
$t_{\mathrm{ls}}$ & complex transmission of the lower sideband&\\
$a_\mathrm{cw}$ & amplitude of the clockwise propagating mode&\\
$a_\mathrm{ccw}$& amplitude of the counterclockwise propagating mode&\\
$\gamma$ & coupling between the two counter propagating modes&\\
$a_+$ & symmetric stationary mode & $a_+ = \left( a_\mathrm{cw} +
a_\mathrm{ccw} \right)/\sqrt{2}$\\
$a_-$ & antisymmetric stationary mode & $a_- = \left( a_\mathrm{cw} -
a_\mathrm{ccw} \right)/\sqrt{2}$\\ 
\end{tabular}
}

%
\newpage

\end{document}